\documentclass[aps,prd,superscriptaddress,showpacs,preprintnumbers,10pt]{revtex4}
\usepackage[dvips]{graphicx,epsfig,color}
\usepackage{wrapfig,rotating}
\usepackage{amssymb,amsmath,array}

\usepackage{color}

\newcommand{\be}{\begin{equation}}
\newcommand{\ee}{\end{equation}}
\newcommand{\bq}{\begin{eqnarray}}
\newcommand{\eq}{\end{eqnarray}}

\newcommand{\D}{\mathrm{d}}
\newcommand{\E}{\mathrm{e}}

\def\Vec#1{\mathpalette{\VVec}{#1}}                  % size-respecting bold
\def\VVec#1#2{\mbox{\boldmath$#1#2$\unboldmath}}

\newcommand{\newangle}{< \hspace{-.8ex} {\scriptscriptstyle )}}

\def\anti#1{\mathpalette{\@anti}{#1}#1}%              % better "anti" bar
\def\@anti#1#2{\sbox0{$#1#2$}%                        % auxiliary for above
  \makebox[0pt][l]{$#1\kern.30\ht0\overline{\kern-.35\ht0\phantom{#2}}$}}

\begin{document}

\title{Extracting the transversity distributions from \\
single-hadron and dihadron production}

\author{Anna~Martin}

\author{Franco~Bradamante}

\affiliation{Dipartimento di Fisica, Universit{\`a}
di Trieste; \\
INFN, Sezione di Trieste, 34127 Trieste, Italy}

\author{Vincenzo~Barone}

\affiliation{Di.S.I.T., Universit{\`a} del Piemonte
Orientale ``A. Avogadro''; \\
INFN, Gruppo Collegato di Alessandria,  15121 Alessandria, Italy}

\begin{abstract}
We present a point-by-point determination of the valence transversity 
distributions from two different types of processes: single-hadron production 
and dihadron production, both in semi-inclusive deep inelastic scattering 
and $e^+ e^-$ annihilation. The extraction is based on some simple
 assumptions and does not require any 
parametrization. The transversity distributions obtained from Collins effect in 
single-hadron production
and from interference effects in dihadron production are found to be 
compatible with each other.  
\end{abstract} 

\pacs{13.88.+e, 13.60.-r, 13.66.Bc, 13.85.Ni}

\maketitle

\section{Introduction}

The transversity distribution, usually called $h_1$, 
is a leading-twist distribution function 
that measures the transverse polarization
of quarks
inside a transversely polarized nucleon (for reviews, see \cite{Barone:2001sp,bbm}). 
Introduced already 35 years ago by Ralston and Soper \cite{Ralston:1979ys}, 
its importance was recognized only recently \cite{Jaffe:1991kp}. It is related to 
the tensor charge of the nucleon and its difference from the 
helicity distribution quantifies the relativistic effects 
in the hadronic structure. 

Being chirally odd, 
transversity decouples from deep inelastic scattering 
and in order to detect it one has to look either at 
hadron-hadron scattering or at hadron leptoproduction. 
Drell-Yan lepton-pair
production with
both colliding hadrons transversely polarized 
involves $h_1$ only, 
but no measurements of this process exist yet. 

Single-spin asymmetries clearly related to the transversity distribution function 
have been measured over the past ten years in semi-inclusive deep inelastic 
scattering (SIDIS) on transversely polarized  nucleons, assessing beyond any doubt 
that transversity is not 
zero \cite{Airapetian:2004tw,Airapetian:2010ds,Alekseev:2010rw,Adolph:2012sn,Airapetian:2008sk,Adolph:2012nw,Adolph:2014fjw}. 

Two processes involving the transversity distributions have been explored so far.  
The first one is single-hadron production 
in SIDIS with a transversely 
polarized target. In this process, transversity 
couples to 
a transverse-momentum dependent chirally odd fragmentation function, $H_1^{\perp}$, 
the so-called 
Collins function \cite{Collins:1993}. This function is independently probed 
in hadron pair production in $e^+ e^-$ annihilation, where it emerges 
in a particular azimuthal correlation of the final hadrons.

The second process 
probing the transversity distributions 
is dihadron production 
in SIDIS with 
a transversely polarized target \cite{Collins:1994ax,Jaffe:1997hf}. In this case, 
transversity couples to 
a dihadron fragmentation function, $H_1^{\newangle}$, and  
the main advantage  
is that it appears collinearly, without involving  
any transverse momentum \cite{Bianconi:1999cd}. 
Again, the fragmentation function $H_1^{\newangle}$ 
can be independently obtained from dihadron pair production in 
$e^+ e^-$ annihilation. 

Recent work points to a close relationship between the Collins 
and the dihadron fragmentation functions \cite{Bradamante:2014a,Bradamante:2014b,Adolph:2014fjw}. 
However, in the present paper the two processes will be treated 
independently as it has been done so far.

Two phenomenological collaborations have extracted the 
valence $u$ and $d$ transversity 
distributions by fitting the SIDIS and the $e^+ e^-$ annihilation asymmetries. 
The Torino group 
\cite{Anselmino:2007fs,Anselmino:2008jk,Anselmino:2013vqa} has used 
only single-hadron data, while the Pavia group \cite{Bacchetta:2011ip,Bacchetta:2013dih} 
has used only dihadron data. The fits of these groups are based on 
different hypotheses, but their results are compatible within 
the present uncertainties.   

In this paper we will follow a different approach. Our aim is to 
extract the transversity point by point in $x$ both from single-hadron 
and dihadron data. 
In particular, we will determine from the $e^+ e^-$ measurements 
the analyzing power of transversely polarized single-hadron and dihadron production,  
and then use this information 
to get the transversity distributions
from the SIDIS data. 
Since we need the widest set of SIDIS observables 
to extract $h_1^{u_v}$ and $h_1^{d_v}$ at the same $x$ and $Q^2$ values, we use 
the COMPASS data, which are available both for a proton and a deuteron 
 target \cite{Ageev:2006da,Adolph:2012sn,Adolph:2012nw,Adolph:2014fjw}. 
As for $e^+ e^-$ scattering, 
we use the Belle data \cite{Abe:2005zx,Seidl:2008xc,Vossen:2011fk}.  
Recently the BaBar Collaboration has also provided data 
on the Collins asymmetry \cite{TheBABAR:2013yha},  but the corresponding data 
for the dihadron production are not yet available.

The main advantage of our approach is that we can analyze
single-hadron and dihadron data in a similar way, 
keeping all sources of uncertainty  
under control and easily checking the robustness of the results.  
We will not need to introduce 
any parametrization of the data. This considerably simplifies 
the determination of the transversity, but, on the other hand,   
prevents us from evolving the distributions. 

The use of consistent 
assumptions for the two sets of processes (single-hadron and dihadron) allows 
a direct comparison of the transversity distributions obtained 
in the two cases.
As we will see, the valence transversities obtained from 
the transverse-momentum-dependent factorization of 
single-hadron production and from the collinear factorization 
of dihadron production turn out to be mutually compatible. This 
is also a check of the validity of the two types of factorization.

The plan of the paper is the following. We start, 
in Sec.~II, from dihadron asymmetries, which allow a simpler analysis. 
In Sec.~III we carry out a similar analysis for the 
Collins asymmetries in single-hadron production and compare the results with those 
obtained in the dihadron case. Sec.~IV contains a general discussion and some concluding 
remarks.

%%%%%%%%%%%%%%%%%%%%%%%%%%%%%%%%%%%%%%%%%%%%%%%%%%%%%%%%%%%%%%%%%%%%%%%%%
\section{Dihadron asymmetries}
\label{twoparticle}
%%%%%%%%%%%%%%%%%%%%%%%%%%%%%%%%%%%%%%%%%%%%%%%%%%%%%%%%%%%%%%%%%%%%%%%%%%%%%

\subsection{Dihadron asymmetries in SIDIS}

The transversity distributions can be probed in 
dihadron leptoproduction from a transversely polarized target, 
$  \ell \,  N^{\uparrow} \,  \rightarrow
 \,  \ell'  \, (h_1 h_2) \,   X $,  
with the two spinless hadrons (typically pions)
in the current jet \cite{Collins:1994ax,Jaffe:1997hf,Bianconi:1999cd}. 
 The idea is to look at an angular correlation 
between the spin of the fragmenting quark and the relative transverse momentum 
of the hadron pair. 
Integrating over the total transverse momentum of the final hadrons, 
one gets a transverse target spin asymmetry in the azimuthal 
angle between a well-defined two-hadron plane and the scattering plane.  
This asymmetry couples the transversity to the dihadron fragmentation function 
$H_1^{\newangle}$ and does not involve the quark transverse momentum. $H_1^{\newangle}$, which 
 originates from the interference of 
different channels in the dihadron formation, describes the fragmentation 
of a transversely polarized quark into a pair of  
spinless hadrons.

The kinematics of the process 
in the $\gamma^* N$ frame involves the following variables:   
the total momentum of the hadron pair $P_{hh} = P_1 + P_2$ 
(with invariant mass $M_{hh}^2 = P_{hh}^2$), the relative momentum 
$R = (P_1 - P_2)/2$,  and the light-cone variables $z_{hh} = P_{hh}^-/\kappa^-$ (which is the fraction 
of the longitudinal momentum $\kappa^-$ of the fragmenting quark carried 
by the hadron pair)  
and $\zeta = 2 R^-/P_{hh}^-$ (which describes how the total longitudinal momentum of the 
pair is split into the two hadrons). 
$\Vec R_T$ is the transverse component of $R$ 
with respect to $\Vec P_{hh}$, and $\phi_R$ is the azimuthal 
angle of $\Vec R_T$ in the plane orthogonal to the $\gamma^*N$ 
axis, measured with respect to the scattering plane. The azimuthal 
angle of the target spin vector is $\phi_S$. Notice that the COMPASS collaboration 
uses, instead of $\Vec R_T$, the vector $\Vec R_{\perp} = (z_2 \Vec P_{1 \perp} 
- z_1 \Vec P_{2 \perp})/(z_1 + z_2)$, which is perpendicular 
to the $\gamma^* N$ axis. The azimuthal angle 
of this vector, which is the relevant variable  
for selecting the asymmetry we are interested in, is the same as $\phi_R$. 
A discussion on the different definitions of the dihadron azimuthal 
angle can be found in Ref.~\cite{Gliske:2014}.

The term in the cross section containing the transversity 
is characterized by the angular modulation 
$\sin (\phi_R + \phi_S)$ and the corresponding measured 
asymmetry is 
\be
A^{hh} (x, z_{hh}, M_{hh}^2, Q^2) 
= 
\, \frac{\sum_{q, \bar q} e_q^2 \,x \,  h_1^q (x, Q^2) 
\, \frac{\vert \Vec R_T \vert}{M_{hh}} \, 
H_{1q}^{\newangle}(z_{hh}, M_{hh}^2, Q^2)}{\sum_{q,\bar q} e_q^2 \,x \, f_1^q (x, Q^2) 
\, D_{1q}^{hh} (z_{hh}, M_{hh}^2, Q^2)}\,, 
\label{twoasym3}
\ee
where $f_1(x, Q^2)$ is the unpolarized distribution function and 
$D_1^{hh} (z_{hh}, M_{hh}^2, Q^2)$ is the unpolarized dihadron fragmentation function.

We now incorporate $\vert \Vec R_T \vert/M_{hh}$ 
into $H_1^{\newangle}$ and integrate over $z_{hh}$ and $M_{hh}^2$:  
\bq
& & 
\widetilde{D}_{1q}^{hh}(Q^2) = \int \D z_{hh} \int \D M_{hh}^2 \, D_{1q}^{hh}(z_{hh}, M_{hh}^2, Q^2) \,,
\label{D1int} \\
& & \widetilde{H}_{1q}^{\newangle}(Q^2) = \int \D z_{hh} \int \D M_{hh}^2 \, 
  \frac{\vert \Vec R_T \vert}{M_{hh}} 
\,  H_{1q}^{\newangle}(z_{hh}, M_{hh}^2, Q^2) \,. 
\label{H1int} 
\eq
The asymmetry can thus be written as 
\be
A^{hh} (x, Q^2) 
=  \frac{\sum_{q, \bar q} e_q^2 \,x \,  h_1^q (x, Q^2) 
\widetilde{H}_{1q}^{\newangle}(Q^2)}{\sum_{q, \bar q} e_q^2 \,x \, f_1^q (x, Q^2) 
\, \widetilde{D}_{1q}^{hh} (Q^2)}\,, 
\label{twoasym3bis}
\ee
Isospin symmetry and charge conjugation
suggest the following relations~\cite{Bacchetta:2011ip,Courtoy:2012ry}: 
\bq
& & \widetilde{D}_{1u}^{hh} = \widetilde{D}_{1d}^{hh} 
= \widetilde{D}_{1 \bar u}^{hh} = \widetilde{D}_{1 \bar d}^{hh} \,,
\;\;\;\; \widetilde{D}_{1s}^{hh} = \widetilde{D}_{1 \bar s}^{hh} \,, \;\;\;\;
\widetilde{D}_{1c}^{hh} = \widetilde{D}_{1 \bar c}^{hh} \,,  
\label{ass2} \\
& & \widetilde{H}_{1u}^{\newangle} = - \widetilde{H}_{1d}^{\newangle} 
= - \widetilde{H}_{1 \bar u}^{\newangle} = \widetilde{H}_{1 \bar d}^{\newangle} \,, 
\;\;\;\;
\widetilde{H}_{1 s}^{\newangle} = - \widetilde{H}_{1 \bar s}^{\newangle} 
= \widetilde{H}_{1 c}^{\newangle} = - \widetilde{H}_{1 \bar c}^{\newangle} =0\,. 
\label{ass1} 
\eq
We also set $\widetilde{D}_{1s}^{hh} = \lambda \, \widetilde{D}_{1 u}^{hh}$, where $\lambda$ 
is a numerical factor which is expected to be smaller than unity. In the following 
we will fix it to 0.5 according to the Monte Carlo simulation of Ref.~\cite{Bacchetta:2011ip}. 

Using the relations (\ref{ass2},\ref{ass1}) and neglecting the charm distribution function, 
the asymmetry for the proton target reduces to 
\be
A_{p}^{hh} (x, Q^2) 
= \frac{4 x h_1^{u_v}(x, Q^2) - x h_1^{d_v}(x, Q^2)}{4  x f_1^{u + \bar u}(x, Q^2) 
+ x f_1^{d + \bar d}(x, Q^2) + \lambda x f_1^{s + \bar s}(x, Q^2)} \, 
\frac{\widetilde{H}_{1 u}^{\newangle}(Q^2)}{\widetilde{D}_{1u}^{hh}(Q^2)}\, 
\label{twoasym_p}
\ee
where $f_1^{q + \bar q} \equiv f_1^q + f_1^{\bar q}$ and $h_1^{q_v} = h_1^q - h_1^{\bar q}$.  
For the deuteron target the asymmetry is 
\be
A_{d}^{hh} (x, Q^2) 
=  
\frac{3 x h_1^{u_v}(x, Q^2) + 3 x h_1^{d_v}(x, Q^2)}{5 x f_1^{u + \bar u}(x, Q^2) 
+ 5 x f_1^{d + \bar d}(x, Q^2)
+ 2 \lambda x f_1^{\bar s + \bar s} (x, Q^2)} \, 
\frac{\widetilde{H}_{1 u}^{\newangle}(Q^2)}{\widetilde{D}_{1u}^{hh}(Q^2)}\,.  
\label{twoasym_d}
\ee

From Eqs.~(\ref{twoasym_p}) and (\ref{twoasym_d}) one can obtain the following combinations 
of the valence transversity distributions (for simplicity we omit the 
dependence on $x$ and $Q^2$) \cite{Bacchetta:2011ip,Bacchetta:2013dih,Elia:2012}
\begin{eqnarray}
& & 4 xh^{u_v}_1 - x h_1^{d_v} =
\frac{1}{\widetilde{a}^{hh}_P}   
 \left (4  x f_1^{u + \bar u} +  x f_1^{d + \bar d} + \lambda  x f_1^{s + \bar s} \right )  A^{hh}_{p}, 
\label{u-d} \\
& & x h_1^{u_v} + x h^{d_v}_1 =
\frac{1}{3} \frac{1}{\widetilde{a}^{hh}_P}  
\left  (5  x f_1^{u + \bar u} + 5  x f_1^{d + \bar d} + 2 \lambda x   f_1^{s + \bar s} \right ) A^{hh}_{d},  
\label{u+d}
\end{eqnarray}
where the analyzing power $\widetilde{a}^{hh}_P$ of the process is defined as 
\be
\widetilde{a}^{hh}_P(Q^2) = 
\frac{\widetilde{H}_{1u}^{\newangle}(Q^2)}{\widetilde{D}_{1 u}^{hh}(Q^2)}. 
\ee 
By combining 
the proton and the deuteron asymmetries one can extract 
the transversity distributions for each flavor: 
\begin{eqnarray}
& & xh^{u_v}_1=
\frac{1}{15} \frac{1}{\widetilde{a}^{hh}_P}  \left [ 
3  \left (4 x f_1^{u + \bar u} + x f_1^{d + \bar d} + \lambda x f_1^{s + \bar s} \right )  A^{hh}_{p} +  
\left  (5 x f_1^{u + \bar u} + 5 x f_1^{d + \bar d} + 2 \lambda x f_1^{s + \bar s} \right ) A^{hh}_{d} 
\right ], 
\label{huv} \\
& & xh^{d_v}_1 =
\frac{1}{15} \frac{1}{\widetilde{a}^{hh}_P}  \left [ - 3 
\left (4 x f_1^{u + \bar u} + x f_1^{d + \bar d} + \lambda x f_1^{s + \bar s} \right )  A^{hh}_{p} +  
4 \left (5  x f_1^{u + \bar u} + 5  x f_1^{d + \bar d} + 2 \lambda x f_1^{s + \bar s} \right ) A^{hh}_{d} 
\right ].  
\label{hdv}
\end{eqnarray}

\subsection{Dihadron asymmetries in $e^+ e^-$ annihilation}

Following the original treatment of Ref. \cite{Bacchetta:2011ip}
we  will extract the analyzing power $\widetilde{a}_P^{hh}$ 
from the measurement of the production of dihadron pairs 
in electron-positron annihilation: $e^+ \, e^- \, \rightarrow \,  (h_1 h_2)\, (h_1' h_2')\, X$, 
where the particles in brackets belong to two back-to-back jets.  
The relevant quantity is the angular correlation of the production planes, 
expressed by the so-called Artru-Collins asymmetry \cite{Artru:1995zu,Boer:2003}.  
The kinematics of the process is described by doubling the 
variables previously introduced. All the variables related 
to the second jet (initiated by the antiquark) will be denoted by a bar. 

If we call $\phi_R$ and $\overline{\phi}_R$ the azimuthal angles of the 
transverse relative momenta $\Vec R_T$ and $\overline{\Vec R}_T$ 
of the two dihadrons, the Artru-Collins azimuthal asymmetry 
is the amplitude of the $\cos (\phi_R + \overline{\phi}_R)$ modulation and  
reads~\cite{Bacchetta:2011ip,Vossen:2011fk,Courtoy:2012ry} (this quantity is called 
$a_{12}$ in Ref.~\cite{Vossen:2011fk})
\bq
A_{e^+e^-}^{hh}(z_{hh}, M_{hh}^2, \overline{z}_{hh}, \overline{M}_{hh}^2, Q^2) 
&=& - \frac{\langle \sin^2 \theta_2 \rangle}{\langle 1+  \cos^2 \theta_2 \rangle} \,  
\nonumber \\ 
& & \times \frac{\sum_{q, \bar q} e_q^2 \, 
\frac{\vert \Vec R_T \vert}{M_{hh}}\, 
H_{1q}^{\newangle}(z_{hh}, M_{hh}^2, Q^2) 
\frac{\vert \overline{\Vec R}_T \vert}{\overline{M}_{hh}}\, 
H_{1 \bar q}^{\newangle}(\overline{z}_{hh}, \overline{M}_{hh}^2, Q^2)}{\sum_{q, \bar q} e_q^2 \, 
D_{1 q}^{hh} (z_{hh}, M_{hh}^2, Q^2) \, D_{1 \bar q}^{hh} (\overline{z}_{hh}, \overline{M}_{hh}^2, Q^2)}
\label{artru}
\eq
where $\theta_2$ is the angle between the $e^+$ and the thrust axis.

Incorporating  $\vert \Vec R_T \vert/M_{hh}$ and 
$\vert \overline{\Vec R}_T \vert/\overline{M}_{hh}$ 
into the interference dihadron fragmentation function and 
integrating over $z_{hh}, \overline{z}_{hh}$ and $M_{hh}^2, \overline{M}_{hh}^2$,  
the Artru-Collins asymmetry becomes
\be
A_{e^+e^-}^{hh} (Q^2)= - \frac{\langle \sin^2 \theta_2 \rangle}{\langle 1+  \cos^2 \theta_2 \rangle} \, 
 \frac{\sum_{q, \bar q} e_q^2 \, 
\widetilde{H}_{1q}^{\newangle}(Q^2) \, 
\widetilde{H}_{1 \bar q}^{\newangle}(Q^2)}{\sum_{q, \bar q} 
e_q^2 \, \widetilde{D}_{1q}^{hh} (Q^2) \, 
\widetilde{D}_{1 \bar q}^{hh} (Q^2)}
\label{artru2}
\ee

We can simplify this expression by using Eqs.~(\ref{ass2}) and (\ref{ass1}) and the relation 
$\widetilde{D}_{1 s}^{hh} = \lambda \widetilde{D}_{1 u}^{hh}$. As for the charm,  
 the Belle experiment \cite{Vossen:2011fk}
finds that 
the $c$ yield is about one half the $uds$, with a 10\% uncertainty.  
Thus we set 
\be
e_c^2 \, \widetilde{D}_{1c}^{hh} \widetilde{D}_{1 \bar c}^{hh} = \mu^2 
\left ( e_u^2 \widetilde{D}_{1 u}^{hh} \widetilde{D}_{1 \bar u}^{hh} + 
e_d^2 \, \widetilde{D}_{1d}^{hh} \widetilde{D}_{1 \bar d}^{hh} + e_s^2 \, 
\widetilde{D}_{1 s}^{hh} \widetilde{D}_{1 \bar s}^{hh} 
\right )\,,    
\ee
with $\mu^2 = 0.5$.  

Finally we get 
\bq
& &   \sum_{q, \bar q} e_q^2 \, \widetilde{H}_{1 q}^{\newangle} 
\widetilde{H}_{1 \bar q}^{\newangle} = - \frac{10}{9} \, 
\left (\widetilde{H}_{1 u}^{\newangle} \right )^2,  
\label{ass3} \\
& &  \sum_{q, \bar q} e_q^2 \, \widetilde{D}_{1 q}^{hh} 
\widetilde{D}_{1 \bar q}^{hh} =  \frac{2}{9} (1 + \mu^2) (5 + \lambda^2)  \, 
\left (\widetilde{D}_{1 u}^{hh} \right )^2 \,.  
\label{ass4}  
\eq
Using eqs.~(\ref{ass3},\ref{ass4}), 
the Artru-Collins asymmetry takes the form
\be
A_{e^+e^-}^{hh} (Q^2)= - \frac{\langle \sin^2 \theta_2 \rangle}{\langle 1+  \cos^2 \theta_2 \rangle} \, 
\frac{5}{(1 + \mu^2) (5 + \lambda^2)} \, 
\left [ \frac{\widetilde{H}_{1u}^{\newangle} (Q^2)}{\widetilde{D}_{1u}^{\newangle} (Q^2)} 
\right ]^2 . 
\label{artru3}
\ee
Solving for the analyzing power, we obtain
\begin{eqnarray}
\vert \widetilde{a}^{hh}_P (Q^2) \vert =  
\left \vert \frac{\widetilde{H}_{1u}^{\newangle}(Q^2)}{\widetilde{D}_{1u}^{hh} (Q^2)} \right \vert = 
\sqrt{-\frac{1}{5} \, (1 + \mu^2) ( 5 + \lambda^2) \frac{\langle 1+  \cos^2 \theta_2 \rangle}
{\langle \sin^2 \theta_2 \rangle} \, A_{e^+ e^-}^{hh} (Q^2)} \,, 
\end{eqnarray}
where the overall sign, which is left undetermined by the $e^+e^-$ data, will be 
chosen in such a way 
to obtain the expected final sign of the transversity distributions, 
that is a positive $h_1^{u_v}$.  
With $\lambda = 0.5$, $\mu^2 = 0.5$ and the numerical values \cite{Vossen:2011fk}
 \begin{eqnarray}
\frac{\langle \sin^2 \theta_2 \rangle}{\langle 1+  \cos^2 \theta_2 \rangle} = 0.7636 \, , \; \; \; \; 
A_{e^+ e^-}^{hh} = - 0.0196 \pm 0.0002 \pm 0.0022
\end{eqnarray}
we find
\begin{eqnarray}
\vert \widetilde{a}^{hh}_P (Q_B^2) \vert = 0.201   
     \;\;\;\; {\rm at} \;\;\;\; Q_B^2 \simeq 110 \; {\rm GeV}^2/c^2
\label{ap_hh}
\end{eqnarray}
with a negligible statistical error and a relative systematic uncertainty of
about 5\%. We also explored another hypothesis for the strange and charm 
contribution to $\widetilde{D}_1^{hh}$, that is 
$\widetilde{D}_{1s}^{hh} = \widetilde{D}_{1c}^{hh}$ 
and $\widetilde{D}_{1s}^{hh} = 0.8 \, \widetilde{D}_{1u}^{hh}$, 
which is also compatible with the Belle finding for the $c$ yield. 
The final result for $\widetilde{a}^{hh}_P$ decreases by 5\%.  

\begin{table}[tb] 
\begin{center}
\begin{tabular}{|c|c|c|c|} 
\hline
               & COMPASS p & COMPASS d & Belle \\ 
\hline 
$\langle z_{hh} \rangle$             & 0.439--0.479   &  0.434--0.485 &  0.4313 \\ 
$\langle M_{hh} \rangle $  
(GeV/c$^2$)       & 0.654--0.723   &  0.633--0.722 &  0.6186 \\ 
\hline
\end{tabular} 
\end{center}
\caption{Mean values of $z_{hh}$ and $M_{hh}$ at COMPASS and Belle.
For COMPASS the ranges refer to the different $x$ bins.} 
\label{tab:kin}
\end{table}

\subsection{Extraction of the transversity distributions}

The value of the analyzing power obtained from $e^+ e^-$ data will be used for the extraction
of the transversity distributions from COMPASS data without any correction, since  
the mean values of $z_{hh}$ and $M_{hh}$ in COMPASS 
are quite close to those of Belle, as shown in Table~\ref{tab:kin}.
Concerning $Q^2$, this is quite different from Belle to COMPASS. 
However, for the purposes of this paper we 
neglect the $Q^2$ evolution
of the analyzing power,  
which has been shown to introduce only a few percent effect \cite{Bacchetta:2011ip}.

\begin{table}[tb] 
\begin{center}
\begin{tabular}{|c|c|rcl|rcl|} 
\hline
$\langle x \rangle$ & $Q^2$ (GeV$^2$/$c^2$) & \multicolumn{3}{c|}{$x h^{u_v}_1$} 
 & \multicolumn{3}{c|}{$x h^{d_v}_1$} \\
\hline
  0.006 & 1.23  &   -0.04 & $\pm$ &  0.04 &   0.08 & $\pm$ &   0.12 \\
  0.010 & 1.48  &    0.03 & $\pm$ &  0.02 &  -0.06 & $\pm$ &   0.07 \\
  0.016 & 1.74  &    0.07 & $\pm$ &  0.02 &   0.17 & $\pm$ &   0.05 \\
  0.025 & 2.09  &    0.02 & $\pm$ &  0.02 &  -0.06 & $\pm$ &  0.05 \\
  0.040 & 2.80  &    0.05 & $\pm$ &  0.02 &   0.00 & $\pm$ &  0.07 \\
  0.062 & 4.34  &    0.06 & $\pm$ &  0.03 &  -0.12 & $\pm$ &  0.09 \\
  0.100 & 6.85  &    0.09 & $\pm$ &  0.04 &  -0.38 & $\pm$ &  0.12 \\
  0.161 & 10.7  &    0.15 & $\pm$ &  0.05 &  -0.30 & $\pm$ &  0.19 \\
  0.280 & 22.0  &    0.25 & $\pm$ &  0.06 &   0.26 & $\pm$ &  0.23 \\
\hline 
\end{tabular} 
\end{center}
\caption{
Values of the valence transversity distributions. Note that the $Q^2$ values 
refer to the proton data. 
 The deuteron data are taken at slightly larger 
$Q^2$,  and in the last bin it is $Q^2=33.2$ GeV$^2$/$c^2$. Errors are statistical only.} 
 
\label{tab:transv}
\end{table} 

\begin{figure}[htb] 
\begin{center} 
\includegraphics[width=0.495\textwidth] {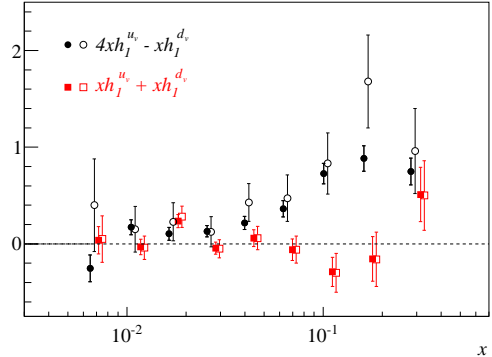}
\end{center}
 \caption{The combinations 
$4 x h_1^{u_v} - x h_1^{d_v}$ (black solid circles) and $x h_1^{u_v} + x h_1^{d_v}$ 
(red solid squares). For comparison 
we plot also the corresponding results of Ref.~\cite{Bacchetta:2013dih} at 
$Q^2 = 2.4$ GeV$^2$ (open points). Squares and open circles are horizontally shifted to make them 
more  visible.}
 \label{fig:comb_hh}
 \end{figure}

Using the dihadron asymmetries measured by COMPASS (the produced hadrons 
are assumed to be pions), the analyzing power (\ref{ap_hh}) extracted 
from Belle data, and the unpolarized PDF's $f_1^{u + \bar u}$ and 
$f_1^{d + \bar d}$ from the CTEQ5D global fit \cite{cteq}, we 
obtain from Eqs.~(\ref{u-d}) and (\ref{u+d})  the combinations 
$4 x h_1^{u_v} - x h_1^{d_v}$ and $x h_1^{u_v} + x h_1^{d_v}$, which 
are shown in Fig.~\ref{fig:comb_hh}.  
The uncertainties shown are statistical only and no uncertainty
on $\widetilde{a}^{hh}_P$ has been considered. We found that the results depend very 
little on the magnitude of the strange contribution, that is on 
the value of $\lambda$. 

In Fig.~\ref{fig:comb_hh} we also plotted the 
same combinations of 
the transversity distributions found by the Pavia group \cite{Bacchetta:2013dih}. 
As one can see, 
the results of our approach are close to those obtained in Ref.~\cite{Bacchetta:2013dih}. 
The differences 
in the combination $4 x h_1^{u_v} - x h_1^{d_v}$, which is 
determined by the proton data, are mainly due to the fact that in 
Ref.~\cite{Bacchetta:2013dih} only the 2007 COMPASS data were used, while 
the present analysis relies on a much wider data set, which includes the 
published 2007 and 2010 data. 

\begin{figure}[htb] 
\begin{center}
\includegraphics[width=0.495\textwidth] {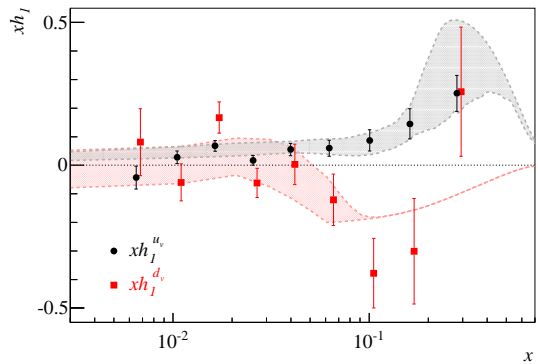}
\end{center}
 \caption{The transversity distributions, $x h_1^{u_v}$ (black circles) and $x h_1^{d_v}$ (red squares) 
extracted from dihadron data.
The shadowed areas are the results of the fits
of the corresponding linear combinations shown in Fig.~\ref{fig:comb_hh}
from Ref.~\cite{Bacchetta:2013dih} at 
$Q^2 = 2.4$ GeV$^2$, in the ``flexible scenario''. For the meaning 
of the bands see Ref.~\cite{Bacchetta:2013dih}.} 
 \label{fig:transv}
 \end{figure}

From Eqs.~(\ref{huv}) and (\ref{hdv}) we get the transversity distributions 
for each flavor separately.   
These are displayed in Fig.~\ref{fig:transv} and their values are collected 
in Table~\ref{tab:transv}, with the corresponding $Q^2$. 
It turns out that  $h_1^{u_v}$ is rather well determined, 
much better than $h_1^{d_v}$, due to the larger uncertainties of the deuteron asymmetry, 
which dominates the $h_1^{d_v}$ extraction, as one can see from Eq.~(\ref{hdv}).
The two distributions have opposite signs, with 
$h_1^{u_v}$ positive 
as a consequence of the sign choice in Eq.~(\ref{ap_hh}). 

For comparison, in Fig.~\ref{fig:transv} we also show the 
results of one of the fits (the so-called ``flexible scenario'') 
of Ref.~\cite{Bacchetta:2013dih} at 
$Q^2 = 2.4$ GeV$^2$/$c^2$ . Again, the agreement between 
our determination and that of the Pavia group is good, 
our points lying within the error bands of their fit. 
In the case of the $d$ quark, the error band of the Pavia fit
at large $x$ shrinks into a line due to the constrain of the Soffer bound.
The broad band for the $u$ quark as compared with the statistical errors of the
present result is very likely due to the
larger uncertainties in the combinations $4 x h_1^{u_v} - x h_1^{d_v}$
as already noted.

\section{Collins asymmetries}

\subsection{Collins asymmetries in SIDIS}

The first evidence of the existence of transversity came from the 
experimental study of 
one-hadron inclusive leptoproduction 
from a transversely polarized proton target, 
$\ell \,   p^{\uparrow} \,   \rightarrow \,  \ell'
\,  h \,  X $,  
at HERMES  \cite{Airapetian:2004tw}. The same signal was subsequently found at 
higher  $Q^2$ by the COMPASS Collaboration 
\cite{Alekseev:2010rw}. 
  
We denote by $P_h$ and $M_h$ the momentum and the mass, respectively, of the produced hadron. 
Conventionally, all azimuthal angles are referred to the lepton scattering plane: 
$\phi_h$ is the azimuthal angle of the hadron $h$, $\phi_S$ is 
the azimuthal angle of the nucleon spin vector $\Vec S_{\perp}$.  
The transverse momenta are defined as follows: $\Vec k_T$ is the transverse momentum of the 
quark inside the nucleon,  $\Vec p_T$ is the transverse momentum 
of the hadron with respect to the fragmenting quark, $\Vec P_{h \perp}$ 
is the measurable transverse momentum of the produced hadron 
with respect to the $\gamma^* N$ axis.

The $\sin (\phi_h + \phi_S)$ term in the cross section 
couples the transverse-momentum dependent transversity distribution $h_1 (x, k_T^2, Q^2)$ to the 
Collins function $H_1^{\perp}(z, p_T^2, Q^2)$,  which describes 
the fragmentation of the transversely polarized struck quark into 
a spinless hadron. The corresponding asymmetry is 
\be
A^h(x, z, Q^2) = 
\frac{\int \D^2 \Vec P_{h \perp} \, 
\mathcal{C} \left [ \frac{\Vec P_{h \perp} \cdot \Vec p_T}{z M_h P_{h \perp}} 
\, h_1 \, H_1^{\perp} \right ]}{\int \D^2 \Vec P_{h \perp} \, 
\mathcal{C} \, \left [ f_1 D_1 \right ]},  
\label{collins}
\ee 
where 
the convolution $\mathcal{C}$ is defined as 
\bq
\mathcal{C} \, [w f D] &=& 
 \sum_a e_a^2 \, x \, \int \D^2 \Vec k_T \int \D^2 \Vec p_T
\, \delta^2 ( z \Vec k_T + \Vec p_T - \Vec P_{h \perp}) \nonumber \\
& & \times \,  w(\Vec k_T, \Vec p_T) 
\, f^a (x,  k_T^2, Q^2) D^a (z,  p_T^2, Q^2)\,. 
\label{convol} 
\eq

If we use, as it is commonly done,  
a Gaussian Ansatz for the transverse-momentum dependent distributions  
\be
f_1(x, k_T^2, Q^2) = f_1(x, Q^2) \, \frac{\E^{- k_T^2/\langle k_T^2 \rangle}}{\pi 
\langle k_T^2 \rangle}\,, \;\;\;\;
h_1(x, k_T^2, Q^2) = h_1(x, Q^2) \, \frac{\E^{- k_T^2/\langle k_T^2 \rangle}}{\pi 
\langle k_T^2 \rangle}\,,
\label{tmd_gauss1} 
\ee
and fragmentation functions 
\be
D_1^{\perp}(z, p_T^2, Q^2)  = D_1(z, Q^2) \, \frac{\E^{- p_T^2/\langle p_T^2 \rangle}}{\pi 
\langle p_T^2 \rangle}\,, \;\;\; 
H_1^{\perp}(z, p_T^2, Q^2)  = H_1^{\perp}(z, Q^2) \, \frac{\E^{- p_T^2/\langle p_T^2 \rangle}}{\pi 
\langle p_T^2 \rangle}\,, 
\label{tmd_gauss2} 
\ee
the Collins asymmetry (\ref{collins}) becomes \cite{Efremov:2006qm} 
\be
A^h(x,z, Q^2) 
= G(z) \,   \frac{\sum_{q,\bar{q}} e_q^2 x h_1^q(x, Q^2)  H_{1q}^{\perp (1/2)}(z, Q^2)}
{\sum_{q,\bar{q}} e_q^2 x f_1^q(x, Q^2)  D_{1q}(z, Q^2)}
\label{collins2}
\ee
where
\be
G (z) = \frac{1}
{\sqrt{1+z^2 \langle k_T^2 \rangle /\langle p_T^2 \rangle}}\,.  
\ee
The ``half-moment'' of $H_1^{\perp}$ is defined as 
\be
H_1^{\perp (1/2)}(z, Q^2) \equiv \int \D^2  \Vec p_T 
\, \frac{p_T}{z M_h} \,  
\, H_1^{\perp}(z, p_T^2, Q^2) \,,   
\ee
and in the Gaussian model is proportional to $H_1^{\perp}(z, Q^2)$,  as defined 
in Eq.~(\ref{tmd_gauss2}):  
\be
H_1^{\perp (1/2)}(z, Q^2) = \frac{\sqrt{\pi \langle p_T^2 \rangle}}{2 z M_h} \, 
H_1^{\perp}(z, Q^2)\,. 
\ee 

In the following we will set  $G (z)=1$,
assuming $z^2 \langle k_T^2 \rangle /\langle p_T^2 \rangle \ll  1$.
This assumption is expected to be reasonable, especially
at low $z$, where the statistics is higher. 

Being interested in the extraction of the transversity distributions, 
we can integrate over $z$, 
\be
\widetilde{H}_{1}^{\perp (1/2)}(Q^2) = 
\int \D z \, H_{1}^{\perp (1/2)}(z, Q^2)\,, 
\;\;\;\;
 \widetilde{D}_{1}(Q^2) = 
\int \D z \, D_{1}(z, Q^2)\,, 
\ee
and write the integrated asymmetry as   
\be
A^{\pm}(x, Q^2) =
\frac{\sum_{q,\bar{q}} e_q^2 x h_1^q(x, Q^2)  \widetilde{H}_{1q}^{\perp (1/2) \pm}(Q^2)}
{\sum_{q,\bar{q}} e_q^2 x f_1^q(x, Q^2)  \widetilde{D}_{1q}^{\pm} (Q^2)}. 
\label{eq:acollc}
\ee
 where the
superscripts 
$+$ and $-$ denote the asymmetries and the fragmentation functions 
for $\pi^+$ and $\pi^-$ production, respectively.

One usually distinguishes favored and unfavored fragmentation functions, defined as 
\begin{eqnarray}
& & D_{1, {\rm fav}} = D_{1u}^{+} = D_{1d}^{-} = D_{1\bar{u}}^{-} 
= D_{1\bar{d}}^{+} \label{eq:favdis1} \\
& & D_{1, {\rm unf}} = D_{1u}^{-} = D_{1d}^{+} = D_{1\bar{u}}^{+} 
= D_{1\bar{d}}^{-}= D_{1s}^{\pm} = D_{1 \bar s}^{\pm}.   
\label{eq:favdis2}
\end{eqnarray}
The corresponding relations for $H_1^{\perp}$ are 
\begin{eqnarray}
& & H_{1, {\rm fav}}^{\perp} = H_{1u}^{\perp +} = H_{1d}^{\perp -} = H_{1\bar{u}}^{\perp-} 
= H_{1\bar{d}}^{\perp +} \label{eq:favh1} \\
& & H_{1, {\rm unf}}^{\perp} = H_{1u}^{\perp -} = H_{1d}^{\perp +} = H_{1\bar{u}}^{\perp +} 
= H_{1\bar{d}}^{\perp -}.    
\label{eq:favh2}
\end{eqnarray}
We assume $H_{1s}^{\perp} 
= H_{1 \bar s}^{\perp} = 0$, as suggested by the string model 
\cite{Artru:2010},
and we ignore the $c$
components of the distribution functions, which are negligible 
at the $x, Q^2$ values of interest here. 
The denominators of the 
asymmetries $\sum_{q, \bar q} e_q^2 x f_1^q \widetilde{D}_{1q}$,
for a proton and a deuteron target ($p, d$) and for charged pions, 
multiplied by 9, can be rewritten as 
\begin{eqnarray}
\hspace{-0.5cm} & & p, \pi^+: 
\;\;\;\;
 x \,  [ 4 (f^{u}_1 + \widetilde{\beta} f^{\bar{u}}_1) +
( \widetilde{\beta} f^{d}_1 + f^{\bar{d}}_1 )  + \widetilde{\beta} (f_1^s + f_1^{\bar s})] \,
 \widetilde{D}_{1, {\rm fav}} 
\equiv x f_p^+ \, \widetilde{D}_{1, {\rm fav}}, 
\label{p+} \\
\hspace{-0.5cm} & & d, \pi^+: 
\;\;\;\; 
  x \,  [ (4 + \widetilde{\beta}) (f^{u}_1 + f^{d}_1 ) 
+ (1 + 4 \widetilde{\beta}) (f^{\bar{u}}_1 + f^{\bar{d}}_1) + 2 \widetilde{\beta} (f_1^s + f_1^{\bar s})]\,
 \widetilde{D}_{1, {\rm fav}}
\equiv x f_d^+ \, \widetilde{D}_{1, {\rm fav}},  
\label{d+} \\
\hspace{-0.5 cm}
& & p, \pi^-: \;\;\;\;
 x \, [4 (\widetilde{\beta} f^{u}_1 + f^{\bar{u}}_1) +
( f^{d}_1 + \widetilde{\beta} f^{\bar{d}}_1 ) + 
\widetilde{\beta} (f_1^s + f_1^{\bar s})] \, \widetilde{D}_{1, {\rm fav}}
\equiv x f_p^- \, \widetilde{D}_{1, {\rm fav}} ,  
\label{p-} \\
\hspace{-0.5cm} 
& & d, \pi^-: \;\;\;\;
x \, [( 1 + 4 \widetilde{\beta}) (f^{u}_1 + f^{d}_1 ) +
( 4 + \widetilde{\beta}) (f^{\bar{u}}_1 + f^{\bar{d}}_1) + 2 
\widetilde{\beta} (f_1^s + f_1^{\bar s}) ]
\, \widetilde{D}_{1, {\rm fav}} 
\equiv x f_d^- \, \widetilde{D}_{1,  {\rm fav}}, 
\label{d-}
\end{eqnarray}
where
\be
\widetilde{\beta} (Q^2) = \frac{\widetilde{D}_{1, {\rm unf}}(Q^2)}{\widetilde{D}_{1, {\rm fav}} (Q^2)}
\label{betaq}
\ee
can be taken from standard parametrizations of fragmentation 
functions.

Similar expressions (with no strange terms) are obtained for the numerator of Eq.~(\ref{eq:acollc}), 
$\sum_{q, \bar q} e_q^2 x h_1^q \widetilde{H}_{1q}^{\perp (1/2)}$, 
with the replacements
$\widetilde{D}_1 \to \widetilde{H}_1^{\perp}$, $f_1 \to h_1$, and $\widetilde{\beta} \to 
\widetilde{\alpha}$,  where
\be
\widetilde{\alpha}(Q^2) = \frac{\widetilde{H}_{1, {\rm unf}}^{\perp (1/2)}(Q^2)}
{\widetilde{H}_{1, {\rm fav}}^{\perp (1/2)}(Q^2)}
\label{alphaq}
\ee
is unknown 
and will be determined later by means of some assumptions. 

Introducing the analyzing power
\be
\widetilde{a}_P^h (Q^2) = \frac{\widetilde{H}_{1, {\rm fav}}^{\perp (1/2)}(Q^2)}
{\widetilde{D}_{1, {\rm fav}} (Q^2)}\,,
\ee 
we find for the proton target
\begin{eqnarray}
& & A^{+}_{p}=
\widetilde{a}_P^h 
\frac{4 (h^{u}_1 + \widetilde{\alpha} h^{\bar{u}}_1) +
 ( \widetilde{\alpha} h^{d}_1 + h^{\bar{d}}_1 ) }{ f_p^+}, 
\label{asym_p+} \\
& & A^{-}_{p}=
\widetilde{a}_P^h 
 \frac{4  (\widetilde{\alpha} h^{u}_1 + h^{\bar{u}}_1) +
 (h^{d}_1 + \widetilde{\alpha}  h^{\bar{d}}_1 ) }{ f_p^-}, 
\label{asym_p-}
\end{eqnarray} 
and for the deuteron target 
\begin{eqnarray}
& & A^{+}_{d} =
\widetilde{a}_P^h 
 \frac{(4 + \widetilde{\alpha})  (h^{u}_1 + h^{d}_1 )  +
(1 + 4 \widetilde{\alpha})  (h^{\bar{u}}_1 + h^{\bar{d}}_1) }{ f_d^+}, 
\label{asym_d+} \\
& & A^{-}_{d} =
\widetilde{a}_P^h 
\frac{(1 + 4 \widetilde{\alpha})  (h^{u}_1 + h^{d}_1 )  +
(4 + \widetilde{\alpha})  (h^{\bar{u}}_1 + h^{\bar{d}}_1) }{ f_d^-} .
\label{asym_d-}
\end{eqnarray}
The combinations 
\begin{eqnarray}
& &  f_p^+ A^{+}_{p} -  f_p^- A^{-}_{p} =
\widetilde{a}_P^h 
(1 - \widetilde{\alpha}) (4 h^{u_v}_1 - h^{d_v}_1  )
\label{p_difference} \\
& &  f_d^+ A^{+}_{d} -  f_d^- A^{-}_{d} =
\widetilde{a}_P^h 
3 (1 - \widetilde{\alpha}) (h^{u_v}_1 + h^{d_v}_1  )
\label{d_difference}
\end{eqnarray}
select the valence transversity distributions.
From eqs.~(\ref{p_difference}, \ref{d_difference}), we get 
the valence distributions for $u$ and $d$ quarks separately: 
\bq
& & x h_1^{u_v} = \frac{1}{5} \frac{1}{\widetilde{a}_P^h (1 - \widetilde{\alpha})} 
\left [ ( x f_p^+ A_p^+ - x f_p^- A_p^-) + \frac{1}{3}  ( x f_d^+ A_d^+ - x f_d^- A_d^-)  \right ] \,, 
\label{uval_coll}
\\
& & x h_1^{d_v} = \frac{1}{5} \frac{1}{\widetilde{a}_P^h (1 - \widetilde{\alpha})} 
\left [ \frac{4}{3}  ( x f_d^+ A_d^+ - x f_d^- A_d^-) -  (x f_p^+ A_p^+ - x f_p^- A_p^-) \right ]\,.
\label{dval_coll}
\eq

Notice that the other two linearly independent combinations of the proton and deuteron 
asymmetries provide the valence + sea transversity distributions
$h_1^{u + \bar u} =h_1^{u} + h_1^{\bar u}$ and $h_1^{d + \bar d}=h_1^{d} + h_1^{\bar d}$: 
\bq
 & & x h_1^{u + \bar u} = \frac{1}{3} \frac{1}{\widetilde{a}_P^h (1 + \widetilde{\alpha})} 
\left [ ( x f_p^+ A_p^+ + x f_p^- A_p^-) - \frac{1}{5} ( x f_d^+ A_d^+ + x f_d^- A_d^-)  \right ] \,, 
\label{u_coll}
\\
& & x h_1^{d + \bar d} = \frac{1}{3} \frac{1}{\widetilde{a}_P^h (1 + \widetilde{\alpha})} 
\left [ \frac{4}{5}  ( x f_d^+ A_d^+ + x f_d^- A_d^-) - ( x f_p^+ A_p^+ + x f_p^- A_p^-) \right ]\,.  
\label{d_coll}
\eq
By further combining Eqs.~(\ref{uval_coll}) and (\ref{dval_coll}), and 
Eqs.~(\ref{u_coll}) and (\ref{d_coll}), one can isolate the sea distributions: 
\bq
& & x h_1^{\bar u} = \frac{1}{15} \frac{1}{\widetilde{a}_P^h (1 - \widetilde{\alpha}^2)} 
\left [ (1 - 4 \widetilde{\alpha}) \, x f_p^+ A_p^+ + (4 - \widetilde{\alpha}) \, x f_p^- A_p^- 
-  x f_d^+ A_d^+ + \widetilde{\alpha}\,  x f_d^- A_d^- \right ], 
\label{ubar} \\
& & x h_1^{\bar d} = \frac{1}{15} \frac{1}{\widetilde{a}_P^h (1 - \widetilde{\alpha}^2)} 
\left [ (4 \widetilde{\alpha} - 1) \, x f_p^+ A_p^+ - (4 - \widetilde{\alpha})\,  x f_p^- A_p^- 
- 4 \widetilde{\alpha} \,  x f_d^+ A_d^+ + 4 \,  x f_d^- A_d^- \right ].  
\label{dbar}
\eq
The overall sea transversity, $h_1^{\bar u }+h_1^{\bar d}$, is determined by 
the deuteron asymmetries only: 
\be
x h_1^{\bar u } + x h_1^{\bar d} = \frac{1}{15} \frac{1}{\widetilde{a}_P^h (1 - \widetilde{\alpha}^2)} 
\left [ (4 + \widetilde{\alpha}) \, x f_d^- A_d^- - (4 \widetilde{\alpha} + 1)\,  x f_d^+ A_d^+ \right ]. 
\label{ubardbar}
\ee

\subsection{Collins asymmetries in $e^+ e^-$ annihilation}

The analyzing power $\widetilde{a}_P^h$ 
 is obtained  from inclusive two-hadron production in 
electron--positron annihilation, 
$  e^+  \,    e^- \, \rightarrow \,  h_1 \,   h_2 \,   X$,  
with the two hadrons in different hemispheres. 
In this process the Collins effect  
is observed in the combination of the fragmenting processes 
of a quark and an antiquark, resulting   
in the product of two Collins functions 
with an overall modulation of the type $\cos (\phi_1 + \phi_2)$, 
where $\phi_1$ and $\phi_2$ are the azimuthal angles of the 
final hadrons around the quark-antiquark axis (approximated by the thrust axis), with respect 
to the $e^+ e^- \to q \bar q$ scattering plane.

The resulting 
$\cos (\phi_1 + \phi_2)$ asymmetry is given by 
\cite{Boer:2008fr,Abe:2005zx,Seidl:2008xc}
\be
A_{e^+e^-}^h (z_1, z_2, Q^2) = \frac{\langle \sin^2 \theta \rangle}{\langle 1 + \cos^2 \theta \rangle}
\, 
\frac{\sum_q e_q^2  H_{1q}^{\perp (1/2)} (z_1, Q^2) \, 
H_{1 \bar q}^{\perp (1/2)} (z_2, Q^2) }{\sum_q e_q^2 
D_{1q}(z_1, Q^2) D_{1\bar q}(z_2, Q^2) }\,, 
\label{a12}
\ee 
having denoted by $z_1 (z_2)$ the fraction of the light-cone momentum 
of the quark (antiquark) carried by the produced hadron. 

The quantity measured in practice is the difference of the asymmetries 
for unlike-sign (U) and like-sign (L) pion pairs, i.e. 
\be
A_{e^+ e^-}^{\rm UL} (z_1, z_2, Q^2) 
= \frac{\langle \sin^2 \theta \rangle}{\langle 1 + \cos^2 \theta \rangle} 
\, \left [ A_{\rm U} (z_1, z_2, Q^2) - A_{\rm L}(z_1, z_2, Q^2) \right ], 
\label{a12ul} 
\ee 
where $A_{\rm U}$ and $A_{\rm L}$, in terms of the favored and unfavored fragmentation functions 
defined as in Eqs.~(\ref{eq:favdis1}) and (\ref{eq:favdis2}), are explicitly  
given by \cite{Anselmino:2007fs}
\bq
& & A_{\rm U}(z_1,z_2, Q^2) =
\frac{5 \, H_{1, {\rm fav}}^{\perp (1/2)}(z_1, Q^2) H_{1, {\rm fav}}^{\perp (1/2)} (z_2, Q^2) 
+ 5 \, H_{1, {\rm unf}}^{\perp (1/2)}(z_1, Q^2) H_{1, {\rm unf}}^{\perp (1/2)} (z_2, Q^2)} 
{5 \, D_{1, {\rm fav}}(z_1, Q^2) D_{1, {\rm fav}}(z_2, Q^2) + 7 \, D_{1, {\rm unf}}(z_1, Q^2) 
D_{1, {\rm unf}}(z_2, Q^2)} \,  
\label{pu} \\
& & A_{\rm L}(z_1,z_2, Q^2) =
\frac{5 \left [  H_{1, {\rm fav}}^{\perp (1/2)}(z_1, Q^2) H_{1, {\rm unf}}^{\perp (1/2)} (z_2, Q^2) 
%+ 5 \, H_{1, {\rm unf}}^{\perp (1/2)}(z_1) H_{1, {\rm fav}}^{\perp (1/2)} (z_2) 
+ (z_1 \leftrightarrow z_2) \right ] } 
{5 \left [ D_{1, {\rm fav}}(z_1, Q^2) D_{1, {\rm unf}}(z_2, Q^2) 
+ (z_1 \leftrightarrow z_2) \right ]
%+ 5 \, D_{1, {\rm unf}}(z_1) D_{1, {\rm fav}}(z_2) 
+ 2 \,  D_{1, {\rm unf}}(z_1, Q^2) 
D_{1, {\rm unf}}(z_2, Q^2) }\,.  
\label{pl} 
\eq
Note that the Belle data 
are subtracted for charm, hence the $c$ components of 
the fragmentation functions have been ignored. 

Let us now introduce the unfavored-to-favored ratios of fragmentation functions
\be
\alpha(z_i, Q^2) = \frac{H_{1, {\rm unf}}^{\perp (1/2)} (z_i, Q^2)}{H_{1, {\rm fav}}^{\perp (1/2)}(z_i, Q^2)}
\,, 
\;\;\;\;\;
\beta(z_i, Q^2) = \frac{D_{1, {\rm unf}} (z_i, Q^2)}{D_{1, {\rm fav}}(z_i, Q^2)}\,, 
\;\;\;\;z_i = z_1, z_2\,.  
\ee
The corresponding integrated quantities have already been defined in 
Eqs.~(\ref{alphaq}) and (\ref{betaq}).

If we set $z_1 = z_2 \equiv z$, that is we use only the diagonal 
measurements, the UL asymmetry can be put in the form 
\be
A_{e^+ e^-}^{\rm UL} (z, Q^2) 
= \frac{\langle \sin^2 \theta \rangle}{\langle 1 + \cos^2 \theta \rangle} 
\, \left [ \frac{H_{1, {\rm fav}}^{\perp (1/2)}(z, Q^2)}{D_{1, {\rm fav}} (z, Q^2)}
\right ]^2 B(z, Q^2) \, 
\label{a12ul2}
\ee
with
\be
B =
\frac{5 + 5 \alpha^2}{5 + 7 \beta^2} - \frac{5 \alpha}{5 \beta + \beta^2}\,. 
\label{eq:aeecfin1b}
\ee
Solving for the analyzing power, we get 
\be
\vert a_P^h (z, Q^2) \vert   
\equiv
\left \vert \frac{H_{1, {\rm fav}}^{\perp (1/2)}(z, Q^2)}{D_{1, {\rm fav}} (z, Q^2)}
\right \vert 
= \sqrt{A_{e^+ e^-}^{\rm UL} (z, Q^2) \frac{\langle 1 + \cos^2 \theta \rangle}{\langle 
\sin^2 \theta \rangle} \, \frac{1}{B(z, Q^2)} }\,.  
\label{rho1}
\ee

While $\beta(z, Q^2)$ and $\widetilde{\beta} (Q^2)$ can be obtained
from standard parametrizations of unpolarized fragmentation functions, 
the functions $\alpha(z, Q^2)$ and $\widetilde{\alpha} (Q^2)$ are not known. 
We consider two different scenarios for the 
relation between the unfavored and the favored Collins function. 

\begin{itemize} 
 \item 
\underline{Scenario 1}. In this scenario we assume that the unfavored 
Collins function is equal and opposite to the favored one, 
\be
H_{1, {\rm fav}}^{\perp (1/2)} (z, Q^2) = - H_{1, {\rm unf}}^{\perp (1/2)} (z, Q^2) \,, 
\label{sc1}
\ee
that is we set $\alpha(z, Q^2) = -1$. This assumption is suggested 
by the fact that 
the asymmetries for positive and negative pions are found to have 
approximately the same size but an opposite sign. 
The Sch{\"a}fer--Teryaev sum rule \cite{Schafer:1999kn}
also points to the same conclusion if interpreted 
as a relation valid for each hadronic species separately \cite{Bacchetta:2007wc}. 

For the ratio of the integrated 
unfavored to favored Collins function, Eq.~(\ref{alphaq}), 
one also gets $\widetilde{\alpha} (Q^2) = -1$. Notice that, in this case, all proton 
and deuteron asymmetries (\ref{asym_p+}-\ref{asym_d-}) depend on the valence transversity 
only, and the sea transversity is undetermined. 

\item
\underline{Scenario 2}. In this scenario we assume $\alpha(z, Q^2) = - \beta(z, Q^2)$, that is 
\be
 \frac{H_{1, {\rm unf}}^{\perp (1/2)} (z, Q^2)}{H_{1, {\rm fav}}^{\perp (1/2)}(z, Q^2)}
= - 
 \frac{D_{1, {\rm unf}} (z, Q^2)}{D_{1, {\rm fav}}(z, Q^2)}\,, 
\label{sc2}
\ee
or equivalently 
\be
 \frac{H_{1, {\rm fav}}^{\perp (1/2)} (z, Q^2)}{D_{1, {\rm fav}}(z, Q^2)}
= - 
 \frac{H_{1, {\rm unf}}^{\perp (1/2)} (z, Q^2)}{D_{1, {\rm unf}}(z, Q^2)}\,, 
\label{sc2bis}
\ee

\end{itemize}

These two scenarios, which lead to quite different expressions for $B(z, Q^2)$, 
can be regarded as a measure of the uncertainty related to the lack of knowledge of the 
different components of the Collins function. We will see, 
however, that the final results are quite insensitive 
to this choice.

We get $D_{1, {\rm unf}}$ and $D_{1, {\rm fav}}$ 
from the DSS parametrization  of fragmentation functions \cite{deflorian}. In the DSS 
fit, actually, $D_{1u}^+$ is not taken to be equal to $D_{1 \bar d}^+$, 
but their difference is rather small. Thus, we identify $D_{1, {\rm fav}}$ 
with $(D_{1 u}^+ + D_{1 \bar d}^+)/2$ as given by DSS.

\begin{figure}[tb] 
\begin{center} %\setlength{\unitlength}{1.0 mm}
\includegraphics[width=0.495\textwidth]{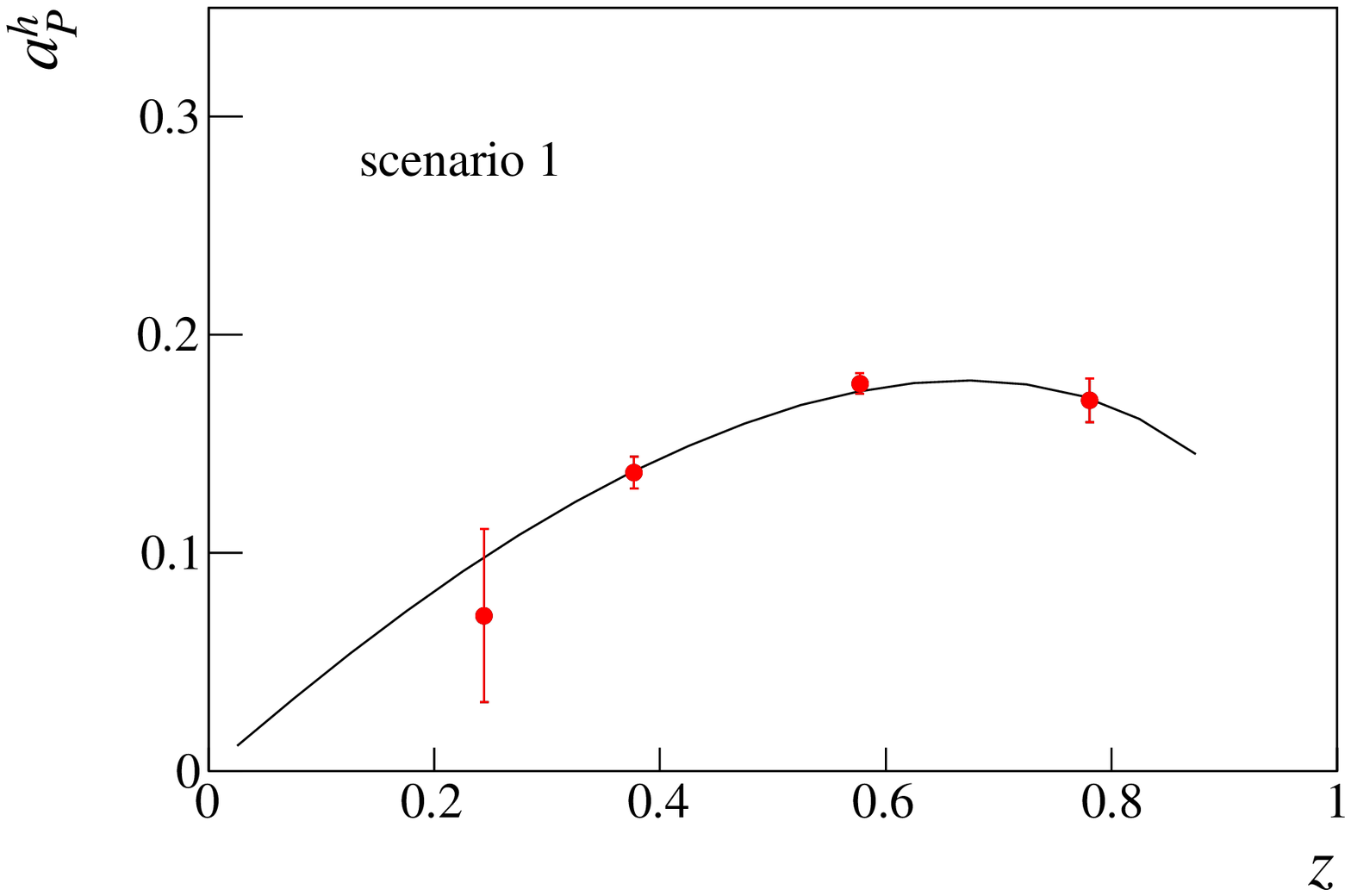}
\includegraphics[width=0.495\textwidth]{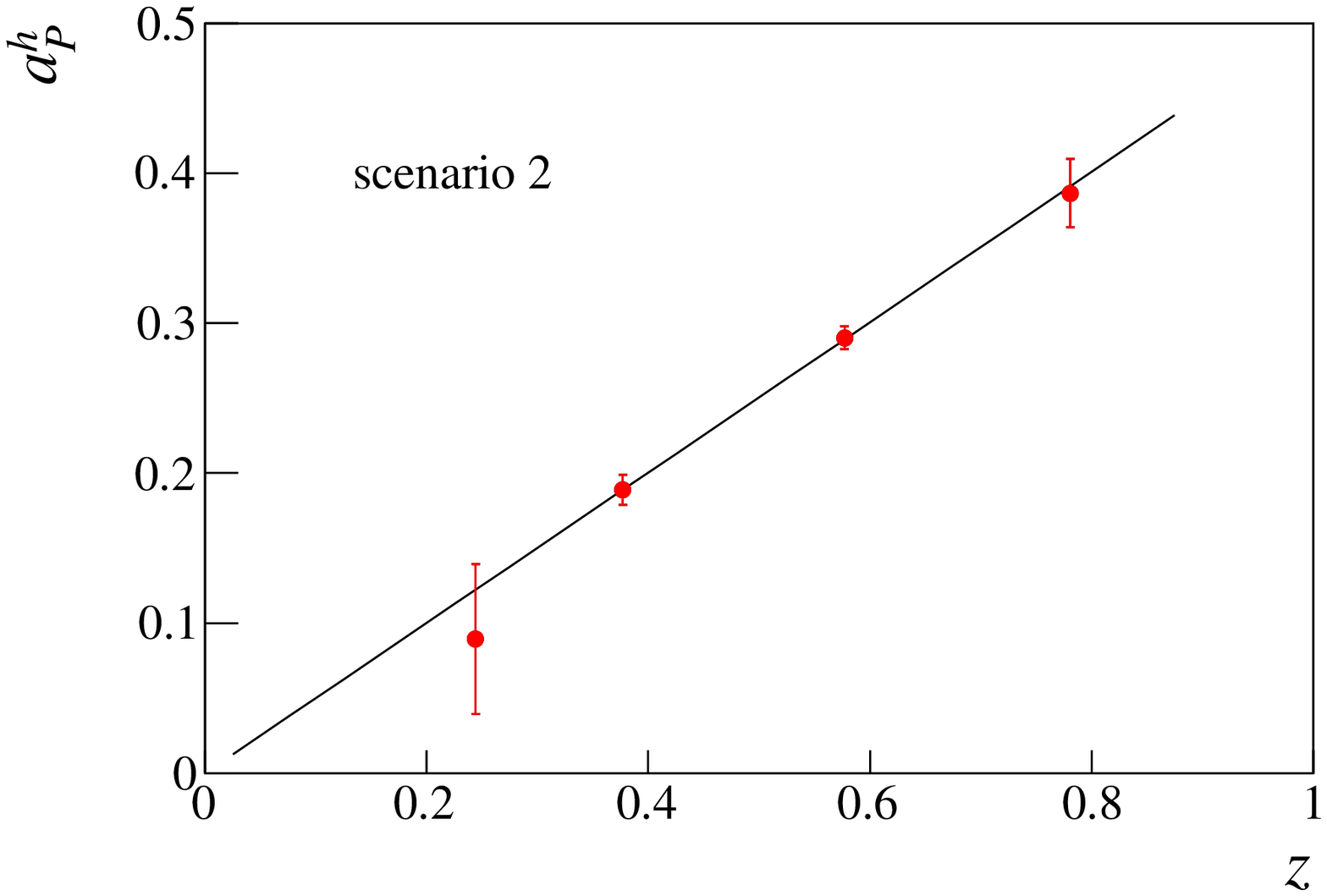}
\end{center}
\vspace*{-.8cm}
 \caption{The analyzing power $a_P^h$ at $Q_B^2 = 110$ GeV$^2$/$c^2$ 
as a function of $z$ for scenario 1 (left) and 2 (right). The curves 
are described in the text.}
% (amextr-1h/apowera1.C)
 \label{fig:apeea1}
 \end{figure}

The experimental values of $A_{e^+ e^-}^{\rm UL}$ at $z_1=z_2=z$
\cite{Seidl:2008xc} are given in Table~\ref{tab:belle1ap}
together with the resulting
analyzing powers in the two scenarios, which are also shown in 
Fig.~\ref{fig:apeea1}. The value $\langle \sin^2 \theta \rangle/\langle 
1 + \cos^2 \theta \rangle = 0.70$ has been used \cite{Seidl:2008xc}.

\begin{table}[tb] 
\begin{center}
\begin{tabular}{|c|c|c|c|} 
\hline
$z$ & $A_{e^+ e^-}^{\rm UL}$  
& $a_P^h (1) $ & $a_P^h (2)$\\
\hline
0.244 & 0.010 $\pm$ 0.011 & 0.071 $\pm$ 0.040 &  0.090 $\pm$ 0.050\\ 
0.377 & 0.046 $\pm$ 0.005 & 0.137 $\pm$ 0.007 & 0.189 $\pm$ 0.010 \\ 
0.577 & 0.113 $\pm$ 0.006 & 0.178 $\pm$ 0.005 & 0.290 $\pm$ 0.008\\ 
0.781 & 0.206 $\pm$ 0.024 & 0.170 $\pm$ 0.010 & 0.387 $\pm$ 0.023\\ 
\hline 
\end{tabular} 
\end{center}
\caption{The $e^+ e^-$ asymmetry measured by Belle \cite{Seidl:2008xc} (with 
statistical and systematic errors added in quadrature) 
and the resulting analyzing powers in scenarios 1 and 2.} 
\label{tab:belle1ap}
\end{table}

Since we need to integrate over $z$ in order to obtain $\widetilde{a}_P^h (Q^2)$,   
we must interpolate in $z$ the values 
of $a_P^h (z, Q^2)$ listed in Table~\ref{tab:belle1ap}.  We use the following 
very simple fitting functions:
\begin{itemize} 
 \item Scenario 1: 
\be
a_P^h (z, Q_B^2)= N z (1- z)^{\gamma}, 
\;\;\;\; Q_B^2 = 110 \; {\rm GeV}^2/c^2\,, 
\ee
with
$N = 0.46 \pm 0.03$ and $\gamma = 0.49 \pm 0.07$.

\item
Scenario 2: 
\be
a_P^h (z, Q_B^2) = N' z \,, 
\;\;\;\;
Q_B^2 = 110 \; {\rm GeV}^2/c^2\,. 
\label{cz}
\ee 
with 
$N' = 0.501 \pm 0.011$. 

\end{itemize} 
These curves are shown in Fig.~\ref{fig:apeea1}.

\subsection{Extraction of the transversity distributions} 

In order to evaluate the analyzing power for the Collins asymmetry  
at the values of $Q^2$ of the COMPASS $x$-bins, we need to evolve the 
fragmentation functions from the Belle value  
$Q_B^2 = 110$ GeV$^2$/$c^2$ to the $Q^2$ values of COMPASS data. 
The evolution of $H_{1}^{\perp (1/2)} (z, Q^2)$ has been worked out 
in Refs.~\cite{Yuan:2009,Kang:2011,Kanazawa:2013}, but involves 
unknown twist-3 fragmentation functions. For our purpose we resort 
to some simplifying assumptions. As a first hypothesis, consistently with what we have 
done in the dihadron case, we assume the analyzing power 
$a_P^h (z, Q^2)$ to be constant in $Q^2$ (another hypothesis 
for the evolution of the Collins function will be discussed later). 
Thus we have  
\be
\widetilde{a}_P^h (Q^2)  = \widetilde{a}_P^h (Q_B^2) 
=\frac{\int \D z \, a_P^h (z, Q_B^2) \, D_{1, {\rm fav}}(z,Q_B^2)} 
{\int \D z \,  D_{1 ,{\rm fav}}(z,Q_B^2)}\,. 
\label{ap_approx_a}
\ee

The values of $\widetilde{a}_P^h$ for 
scenarios 1 and 2 are 
\be
{\rm Scenario} \; 1: \;\;
\widetilde{a}_P^h  = 0.122, 
\;\;\;\;\;
{\rm Scenario} \; 2: \;\;
\widetilde{a}_P^h = 0.173.  
\label{ap12}
\ee

It is now possible to extract the transversity distributions 
from the Collins asymmetries $A^{\pm}_{p/d}$. 
The last ingredient we need is the ratio $\alpha(Q^2)$ 
of the integrated unfavored to favored Collins functions, 
Eq.~(\ref{alphaq}). Whereas for scenario 1 we have 
$\widetilde{\alpha}(Q^2) = -1$, for scenario 2, 
using $\alpha(z, Q^2) = - \beta(z, Q^2)$ and the linear behavior of Eq.~(\ref{cz}), 
we find  
\be
\widetilde{\alpha}(Q^2) = - \frac{\int \D z \, z \, D_{1, {\rm unf}}(z, Q^2)}{
\int \D z \, z \, D_{1, {\rm fav}}(z, Q^2)}\,. 
\label{alphaq_sc2}
\ee
In the COMPASS $Q^2$ range, $\widetilde{\alpha}$
ranges from $-0.43$ at the highest $x$ value to $-0.34$ at the lowest $x$ value. 

\begin{figure}[htb] 
\begin{center} %\setlength{\unitlength}{1.0 mm}
\includegraphics[width=0.495\textwidth]{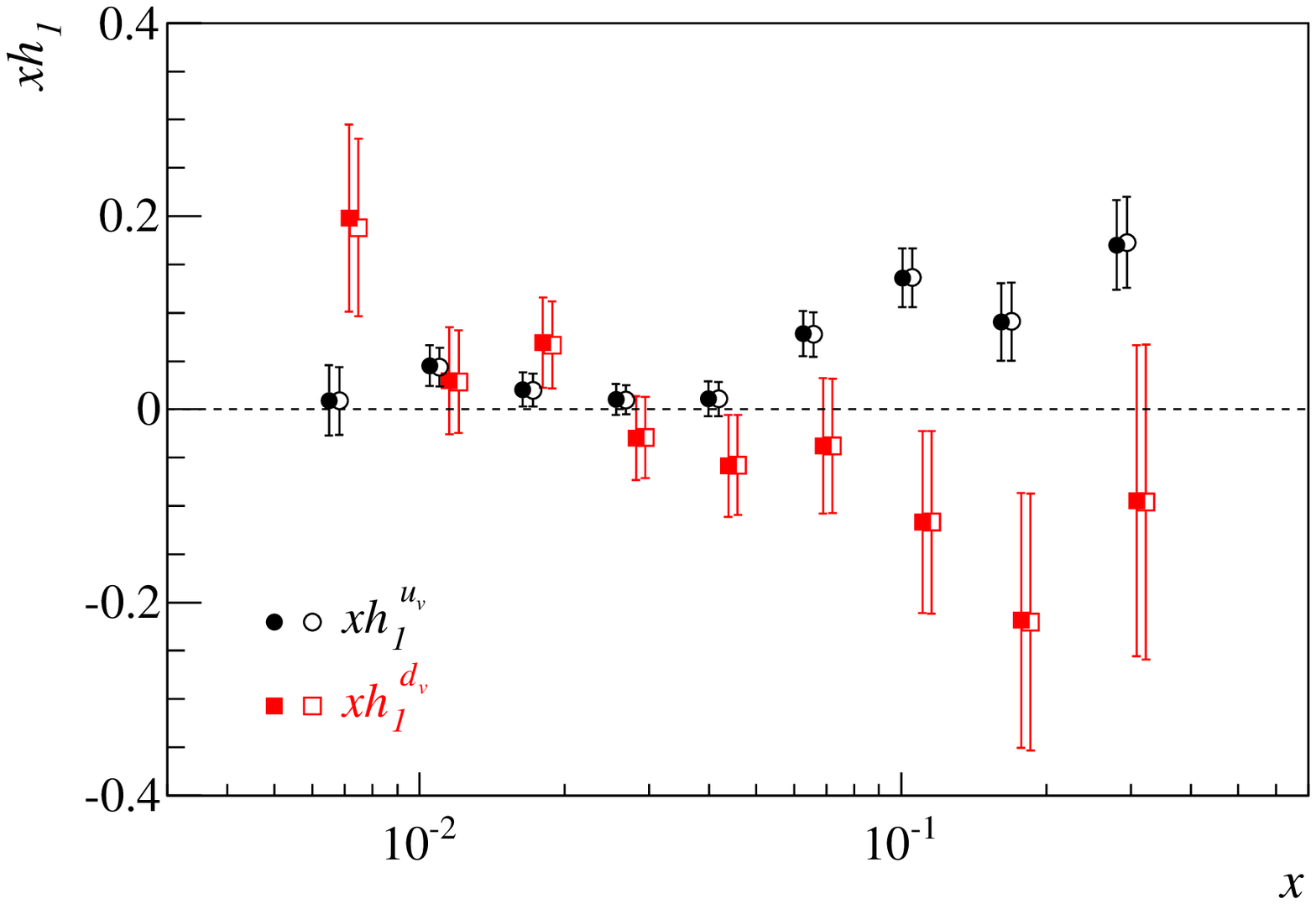}
\includegraphics[width=0.495\textwidth]{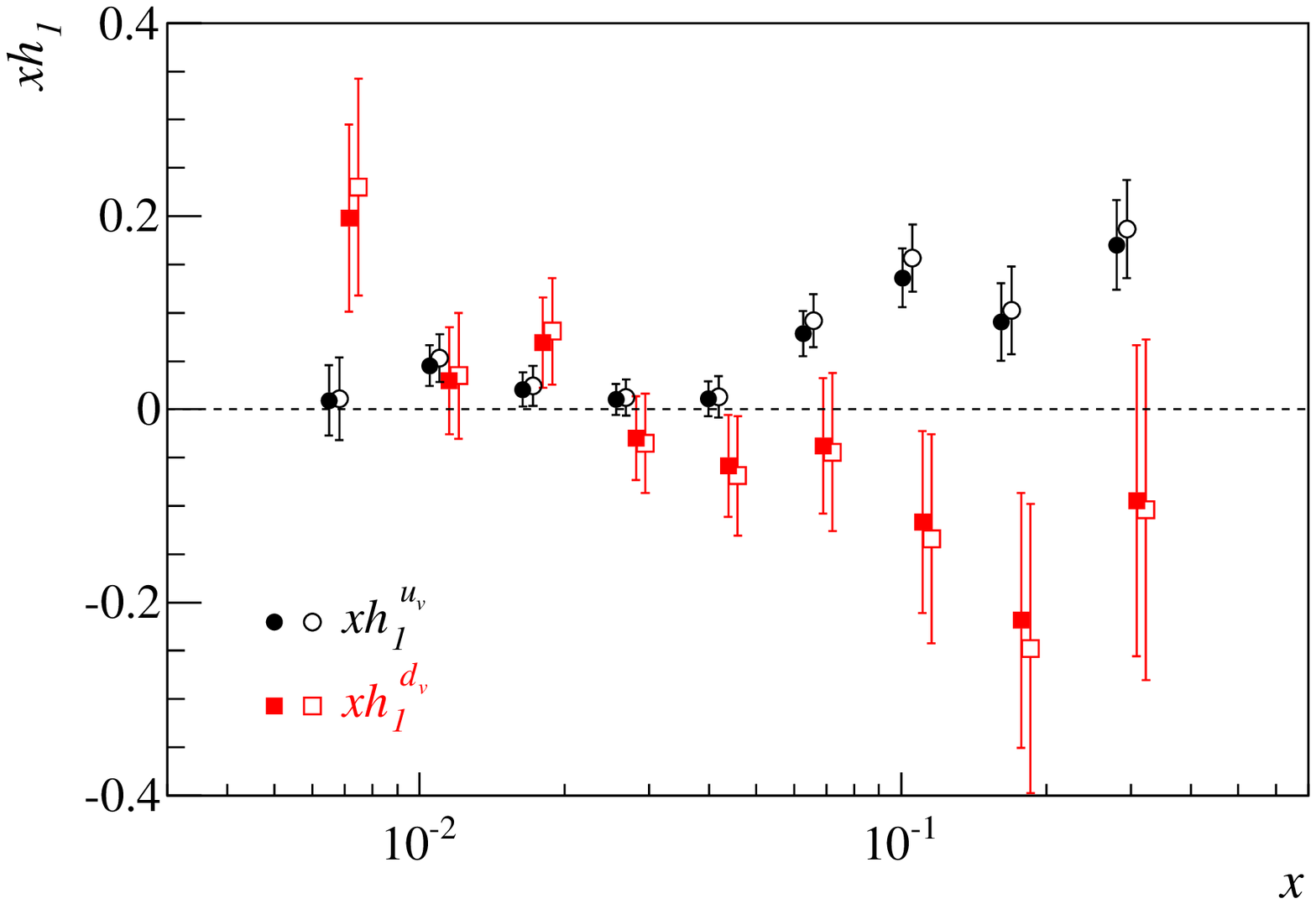}
\end{center}
 \caption{
Left: Valence transversity distributions for Scenario 1 (open points) 
and Scenario 2 (solid points). 
Right: Valence transversity distributions for Scenario 2 
and the default, Eq.~(\ref{ap_approx_a}) (solid points), and alternative, Eq.~(\ref{ap_approx_b}) 
(open points), 
$Q^2$ evolution hypothesis. 
In both plots, black circles represent $x h_1^{u_v}$, 
red squares represent $x h_1^{d_v}$. }
 \label{fig:hud12}
 \end{figure}

Using the CTEQ5D unpolarized distribution functions \cite{cteq} 
and the DSS unpolarized fragmentation functions, and 
inserting the values (\ref{ap12}) and the asymmetries measured by COMPASS
into Eqs.~(\ref{uval_coll}) and (\ref{dval_coll}), we finally find the 
valence transversity distributions plotted in Fig.~\ref{fig:hud12} (left). 
As one can see the distributions for the two Scenarios 
are very close to each other. 
This is due to the fact that 
the different assumptions for the relation between the favored and 
the unfavored Collins functions, leading to a different $\widetilde{\alpha}$, 
 are compensated by the difference in 
the analyzing powers $\widetilde{a}_P^h$ extracted from the $e^+ e^-$ data in the 
two scenarios, so that the product $\widetilde{a}_P^h (1 - \widetilde{\alpha})$, 
which would discriminate between the two scenarios, is actually almost the same.
Very much as in the dihadron case
the valence $u$ quark transversity distribution is definitively positive
and well determined while the $d$ quark has about the same size but
opposite sign and has considerably larger uncertainties. 

To check the robustness of our results also against a different hypothesis 
on the evolution of the fragmentation functions, we now assume that 
$H_{1, {\rm fav}}^{\perp}$ evolves very little compared to $D_{1, {\rm fav}}$, which implies, 
instead of Eq.~(\ref{ap_approx_a}), 
\be
\widetilde{a}_P^h (Q^2) 
=\frac{\int \D z \,  a_P^h (z, Q_B^2) \, D_{1, {\rm fav}}(z,Q_B^2)} 
{\int \D z \,  D_{1, {\rm fav}}(z,Q^2)}\, .
\label{ap_approx_b}
\ee
We find that the resulting transversity distributions are close to those  
obtained under the assumption of Eq.~(\ref{ap_approx_a}),
as can be seen in Fig.~\ref{fig:hud12} (right).
Their difference is of about 10\% at large $x$.

\begin{figure}[htb] 
\begin{center} %\setlength{\unitlength}{1.0 mm}
\includegraphics[width=0.495\textwidth]{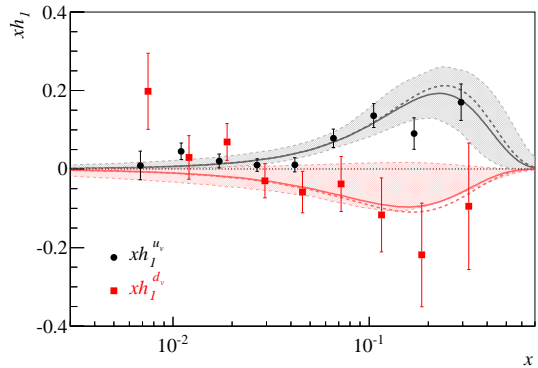}
\end{center}
 \caption{Valence transversity distributions, $x h_1^{u_v}$ (black circles) 
and $x h_1^{d_v}$ (red squares) for Scenario 2, compared to 
the corresponding distributions extracted in Ref.~\cite{Anselmino:2013vqa}
at $Q^2=10$ GeV$^2$/$c^2$ (curves and error bands). The error bands refer to 
$Q^2 = 10$ GeV$^2$/$c^2$. The solid line and the dashed line show
the central values at $Q^2=10$ GeV$^2$/$c^2$ and $Q^2 = 2$ GeV$^2$/$c^2$, respectively. For the 
meaning of the bands see Ref.~\cite{Anselmino:2013vqa}.}  
 \label{fig:hud1_Torino}
 \end{figure}

In Fig.~\ref{fig:hud1_Torino} we compare the results of the present paper with 
the transversity distributions extracted by the Torino group \cite{Anselmino:2013vqa},
at $Q^2=10$ GeV$^2$/$c^2$. The effect of the evolution is however small, 
as
shown by the solid and dashed lines, 
which refer to $Q^2$ = 10 GeV$^2$/$c^2$ and $Q^2$ = 2 GeV$^2$/$c^2$, 
respectively.
Our point-by-point determination turns out to be in good agreement with
the fit of Ref.~\cite{Anselmino:2013vqa} which includes also the HERMES
proton data and all the asymmetries measured as functions of
$z$ and $p_T$.  Recently,  an extraction of the transversity 
using an approximate transverse-momentum dependent evolution at the next-to-leading 
logarithmic order has been performed \cite{Kang:2014}, with 
results very similar to those found here.

\begin{figure}[htb] 
\begin{center} %\setlength{\unitlength}{1.0 mm}
\includegraphics[width=0.495\textwidth]{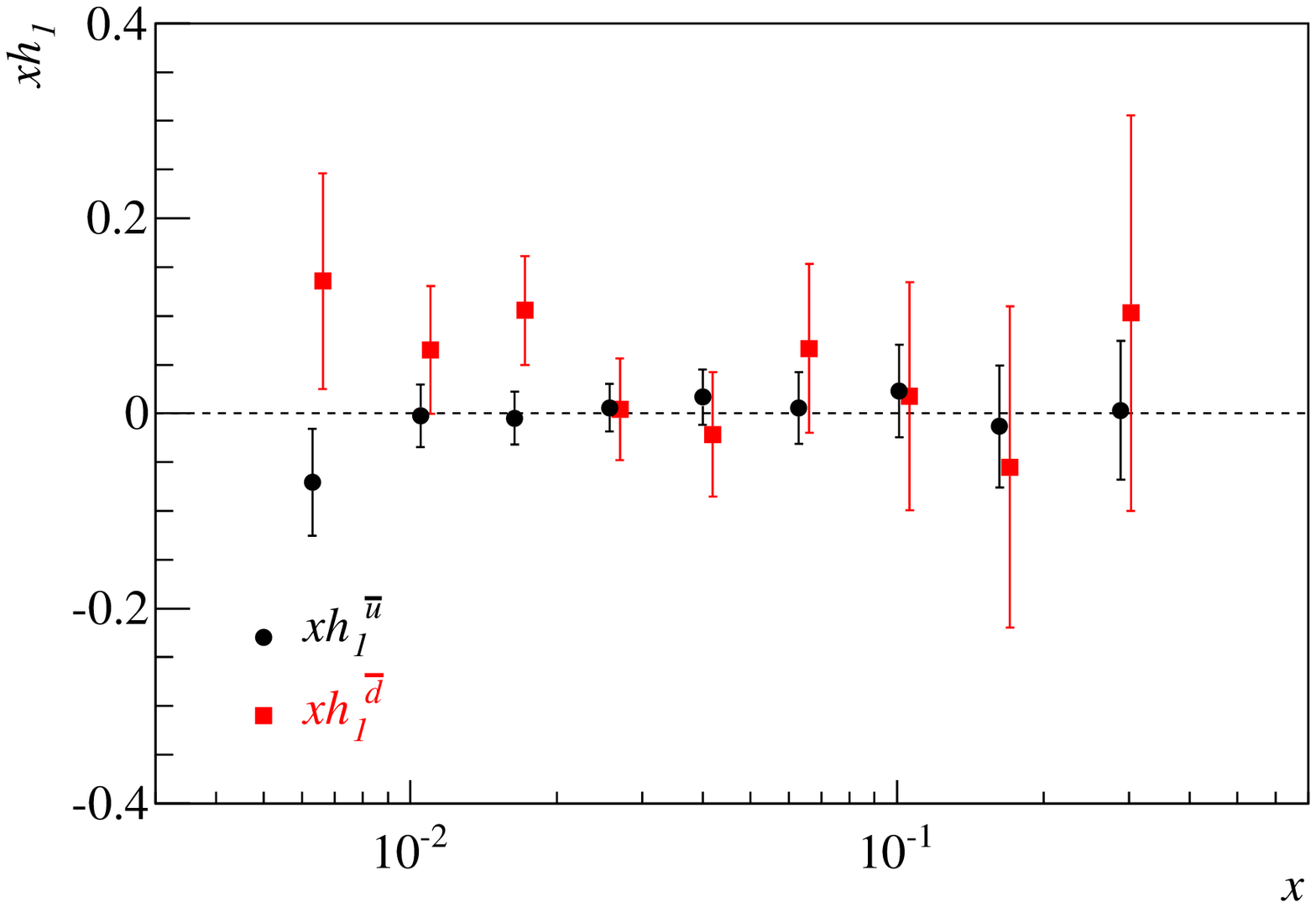}
\includegraphics[width=0.495\textwidth]{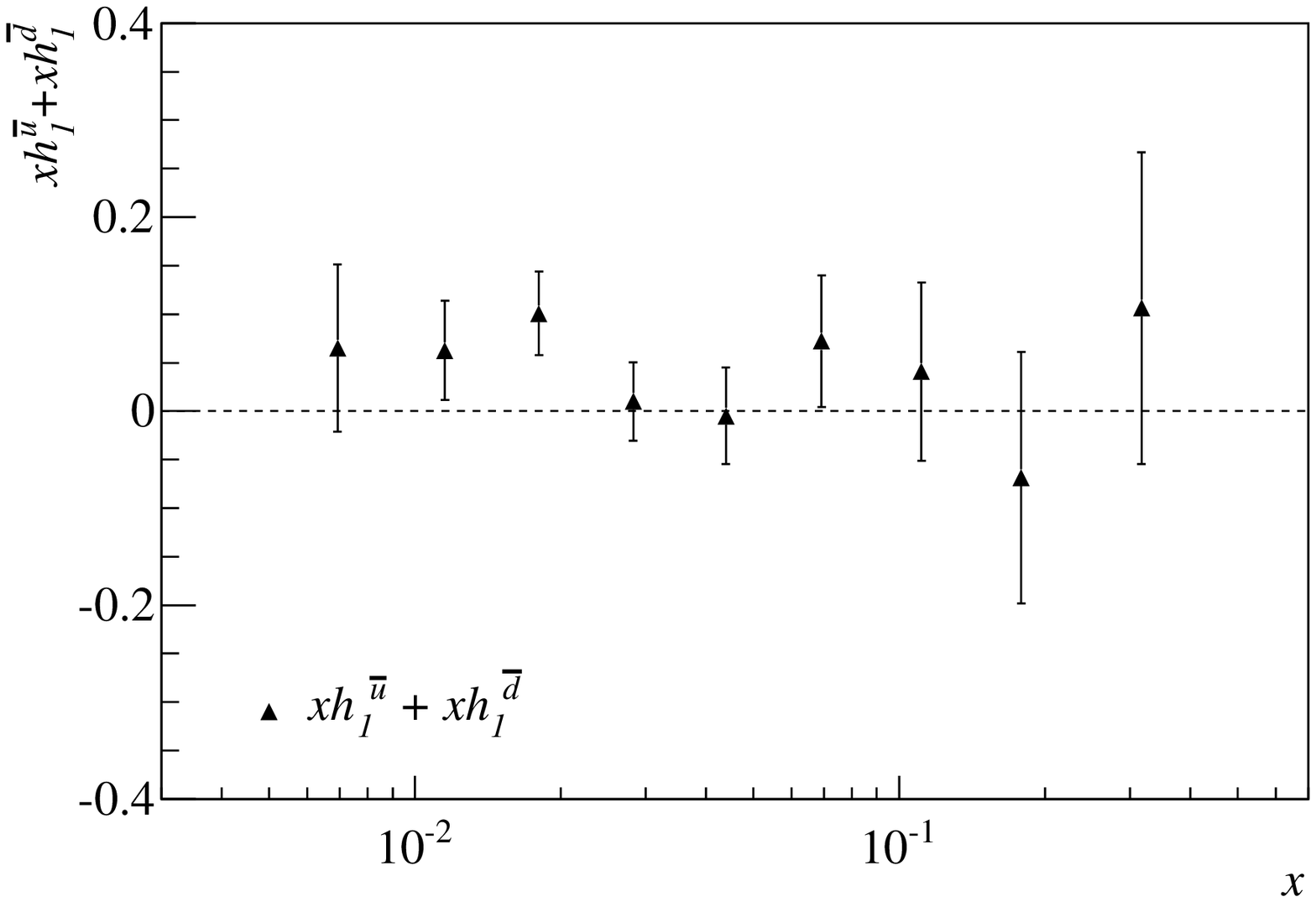}
\end{center}
 \caption{The sea transversity distributions $x h_1^{\bar u}$ and $x h_1^{\bar d}$
(left) and  $x h_1^{\bar u} + x h_1^{\bar d}$ (right) in  Scenario 2.  }
 \label{fig:hsea}
 \end{figure}

In Scenario 2, where $\widetilde{\alpha} \neq -1$, we can use 
Eqs.~(\ref{ubar}-\ref{ubardbar}) to determine the sea transversity distributions. 
This is not possible in Scenario 1, however we regard the fact that the 
two scenarios lead to results which are essentially identical for the valence 
transversity as a justification to our procedure. 
The resulting values for the sea transversity distributions are shown in 
Fig.~\ref{fig:hsea} (left) individually for ${\bar u}$ and ${\bar d}$, 
while the combined 
distribution $x h_1^{\bar u} + x h_1^{\bar d}$, obtained 
from Eq.~(\ref{ubardbar}),  is shown in Fig.~\ref{fig:hsea} 
(right). 
Both the $\bar{u}$ and the $\bar{d}$ values are compatible with zero, but,  
as it is apparent 
from the results, the $\bar{u}$ transversity values have an accuracy a factor of 
3 better than the $\bar{d}$ values. 

The difference in sensitivity between the $u$ and 
the $d$ quark, which affects both the valence and the sea distributions, is 
due to the fact that the COMPASS deuteron data have larger errors than the 
proton data. 
In the particular case of the sea distributions, from Eq.~(\ref{ubardbar}) 
it is clear 
that $x h_1^{\bar u} + x h_1^{\bar d}$ is determined directly from the Collins 
asymmetry of the deuteron data, which have rather large error bars. 
On the contrary, the $\bar{u}$ distribution alone is well determined because in 
the right hand side of Eq.~(\ref{ubar}) the proton data have a considerably 
larger weight than the deuteron data.

All the values of the transversity distributions in the different $x$
bins and the corresponding $Q^2$ are given in Table~\ref{tab:alltr}.
The results from the extraction of transversity from the Collins asymmetries 
are the ones from Scenario 2, but as we have shown the two scenarios we have 
used led to essentially identical results.

\begin{table}[tb] 
\begin{center}
\begin{tabular}{|c|c|rcl|rcl|rcl|rcl|} 
\hline
$\langle x \rangle$ & $Q^2$ (GeV$^2$/$c^2$) & \multicolumn{3}{c|}{$x h^{u_v}_1$}  &
\multicolumn{3}{c|}{$x h^{d_v}_1$} 
& \multicolumn{3}{c|}{$x h^{\bar{u}}_1$}  & \multicolumn{3}{c|}{$x h^{\bar{d}}_1$}  \\
\hline
  0.006 & 1.27  &    0.01 & $\pm$ & 0.04 &  0.23 & $\pm$ & 0.11 & -0.07& $\pm$ & 0.05 &   0.14& $\pm$ & 0.11 \\
  0.010 & 1.55  &    0.05 & $\pm$ & 0.03 &  0.03 & $\pm$ & 0.06 &  0.00& $\pm$ & 0.03 &   0.07& $\pm$ & 0.07 \\
  0.016 & 1.83  &    0.02 & $\pm$ & 0.02 &  0.08 & $\pm$ & 0.06 & -0.01& $\pm$ & 0.03 &   0.11& $\pm$ & 0.06 \\
  0.025 & 2.17  &    0.01 & $\pm$ & 0.02 & -0.03 & $\pm$ & 0.05 &  0.01& $\pm$ & 0.02 &   0.00& $\pm$ & 0.05 \\
  0.040 & 2.83  &    0.01 & $\pm$ & 0.02 & -0.07 & $\pm$ & 0.06 &  0.02& $\pm$ & 0.03 &  -0.02& $\pm$ & 0.06 \\
  0.063 & 4.34  &    0.09 & $\pm$ & 0.03 & -0.04 & $\pm$ & 0.08 &  0.00& $\pm$ & 0.04 &   0.07& $\pm$ & 0.09 \\
  0.101 & 6.76  &    0.16 & $\pm$ & 0.04 & -0.13 & $\pm$ & 0.11 &  0.02& $\pm$ & 0.05 &   0.02& $\pm$ & 0.12 \\
  0.163 & 10.5  &    0.10 & $\pm$ & 0.04 & -0.25 & $\pm$ & 0.15 & -0.01& $\pm$ & 0.06 &  -0.06& $\pm$ & 0.17 \\
  0.288 & 22.6  &    0.19 & $\pm$ & 0.05 & -0.10 & $\pm$ & 0.18 &  0.00& $\pm$ & 0.07 &   0.10& $\pm$ & 0.20 \\
\hline 
\end{tabular} 
\end{center}
\caption{
Values of the valence and sea transversity distributions
from the Collins asymmetries for Scenario 2. 
Note that the $Q^2$ values 
refer to the proton data. The deuteron data are taken at slightly larger 
$Q^2$ and in the last bin it is $Q^2=25.9$ GeV$^2$/$c^2$. Errors are statistical only. 
}
\label{tab:alltr}
\end{table} 

\section{Discussion of the results and concluding remarks}

\begin{figure}[tb] 
\begin{center} %\setlength{\unitlength}{1.0 mm}
\includegraphics[width=0.495\textwidth]{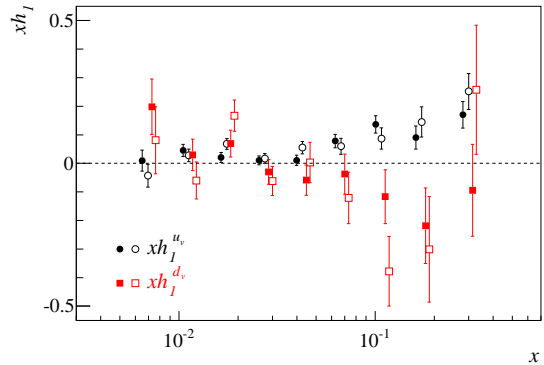}
\end{center}
 \caption{The valence transversity distributions from dihadron (open points) 
and Collins asymmetries (solid points). Black circles represent 
$x h_1^{u_v}$, red squares represent $x h_1^{d_v}$. The transversity 
extracted from single-hadron leptoproduction refers to Scenario 2.}
 \label{fig:hud_comp}
 \end{figure}

We now compare the results for the transversity extracted 
from the dihadron asymmetries and from the Collins asymmetries 
in single--hadron leptoproduction. 
The valence transversity  distributions obtained from these 
two types of processes are shown in 
 Fig.~\ref{fig:hud_comp}. The agreement is quite impressive.

    The error bars are computed from the statistical errors of the measured 
asymmetries as quoted by the experimental Collaborations, and no attempt has 
been made to try to assign a systematic error to the results. For the Collins extraction, 
the fact that different scenarios for the 
$H_{1, {\rm fav}}^{\perp (1/2)}/H_{1, {\rm unf}}^{\perp (1/2)}$ ratio  
and for the evolution lead to results which differ only by few percent is an indication 
that in our approach the systematic uncertainties related to the phenomenological analysis 
are much smaller than the statistical errors. 
The transversity values obtained from the dihadron asymmetries and 
from the Collins asymmetries are very well compatible within themselves, and 
clearly support the fact that the same distributions are measured in the  
two processes. 

Another relevant observation is that there is no indication for a bias due
to the Gaussian ansatz and to the assumption $G=1$ used to extract the 
transversity distributions from the Collins asymmetry.

It is also clear that the $u$ quark transversity is determined with a much 
better accuracy than the $d$ quark transversity, due to the fact that 
the asymmetry measurements on the proton are considerably more accurate than 
the corresponding ones on the deuteron, in particular in the valence region 
(the COMPASS Collaboration has taken data about 7 times less on deuterons 
than on protons). 
This unbalance can only increase if in a global analysis also the HERMES data 
are included, which were taken only on protons. 
Still, within the accuracy of 
the data, the $d$ quark valence transversity distribution is definitively 
different from zero, and of about the same size as the corresponding $u$ quark 
distribution, but with opposite sign.

    Another interesting result from our work is the fact that our procedure 
has allowed also the extraction of the transversity distributions of the
$\bar{u}$ and the $\bar{d}$ quarks. 
They are both found to be compatible with zero, but it is interesting to 
underline that the accuracy of this result in the case of $\bar u$
is quite good, comparable to that 
of the $u$ valence distribution.
 
   To conclude, in a simple and direct model-independent way we have 
extracted the $u$ and $d$ quark transversity distributions, both valence and 
sea, from the COMPASS and the Belle data. 
The method seems robust, and probably can be extended to extract other 
distribution functions, in particular the Sivers and the Boer-Mulders functions.

    To improve on the knowledge of transversity clearly more data are needed, 
in particular on the deuteron. 
The long-term solution to this quest is the planned future Electron Ion 
Collider (EIC), but in the near future measurements at JLAB12 and the 
proposed new COMPASS run on a deuteron target \cite{EPSG} will be highly 
beneficial.

\begin{acknowledgments}
One of us (V.B.) acknowledges the kind hospitality of the 
Dipartimento di Fisica of the University of Trieste, where 
this work was done. We acknowledge partial 
support from the Universit\`a degli Studi di Trieste in the framework 
of ``Finanziamento
di Ateneo per progetti di ricerca - FRA2012''. 
We are grateful to Stefano Melis and to Alessandro Bacchetta 
for providing us the curves of their fits.

\end{acknowledgments}

\end{document}